\documentclass[12pt]{article}
\usepackage{slashed}
\usepackage{amsmath}
\usepackage{amsfonts}
\usepackage{indentfirst}
\usepackage{color}
\newcommand{\nin}{\noindent}
\newcommand{\be}{\begin{equation}}
\newcommand{\ee}{\end{equation}}
\newcommand{\bea}{\begin{eqnarray}}
\newcommand{\eea}{\end{eqnarray}}

\newcommand{\nn}{\nonumber\\}

\newcommand{\ol}{\overline}

\begin{document}

\begin{flushleft}
KCL-PH-TH/2013-{\bf 32} 
\end{flushleft}

\begin{center}
{\bf Lorentz-symmetry violation and dynamical flavour oscillations}

\vspace{0.5cm}

J. Alexandre\\
King's College London, Department of Physics, WC2R 2LS, UK
jean.alexandre@kcl.ac.uk

\vspace{0.5cm}

{\bf Abstract}

\end{center}

\vspace{0.2cm}

We show how a mass mixing matrix can be generated dynamically, for two massless fermion flavours coupled to a Lorentz invariance violating (LIV) gauge field. 
The LIV features play the role of a regulator for the gap equations, and the non-analytic dependence of the dynamical masses, as functions
of the gauge coupling, allows
to consider the limit where the LIV gauge field eventually decouples from the fermions. 
Lorentz invariance is then recovered, to describe the 
oscillation between two free fermion flavours, and we check that the finite dynamical masses are the only effects of the original LIV theory.

\section{Introduction}

The generation of quark, lepton and vector boson masses, as described in the Standard Model due to their coupling with the 
Higgs boson, 
seems to have been confirmed by the latest experimental results at the Large Hadron Collider \cite{Higgs}, with the discovery
of a Higgs-like (scalar) particle. 
However, the origin of neutrino masses is still not well established, although the seesaw mechanism seems the most elegant and 
simple for such a purpose~\cite{seesaw}. 

The possibility, therefore, 
of generating neutrino masses dynamically is still at play. 
A scenario has been proposed in \cite{ALM}, in which flavour oscillations can arise dynamically from the flavour-mixing interaction
of two massless bare fermions with an Abelian gauge field, which has a Lorentz-Invariance-Violating (LIV) propagator.
Lorentz symmetry violation is achieved by higher order space derivatives, which are suppressed by a large mass scale $M$.
This mass scale allows the dynamical generation of fermion masses, as was shown in \cite{JA} with the Schwinger-Dyson approach.
Another role of this mass scale is to lead to a finite gap equation, and therefore to regulate the model.

Moreover, LIV U(1) gauge models of the form suggested in \cite{JA} have been shown to arise in the low-energy limit of some 
consistent quantum gravity theories \cite{NM}, for instance when the U(1) gauge theory is embedded in a stringy space time foam model, with the 
foamy structures being provided by (point-like) D-brane space-time defects (``D-particles''). In such microscopic models, the gauge field 
is one of the physical excitations on brane world universes interacting with the D-particles.
It was observed in \cite{NM} that the LIV Lagrangian of \cite{JA} can be obtained from a Born-Infeld-type Lagrangian of the U(1) gauge field 
in the D-particle background, 
upon an expansion in derivatives. Lorentz Violation arises locally in such models as a result of the recoil of the D-particle defects during 
their interaction with open strings representing the U(1) excitations. Finally, such a microscopic model has been used for the study of decoherence 
in neutrino oscillations \cite{AFMP}.

An important point is the following structure of the fermion dynamical mass \cite{JA}
\be\label{mdyn}
m_{dyn}\simeq M \, \exp(-a/e^2)~, 
\ee
where $a$ is a positive constant and $e$ is the gauge coupling. 
From the expression (\ref{mdyn}), one can see that it is possible to take the simultaneous limits
$M\to\infty \quad {\rm and} \quad e \to 0$, 
in such a way that the dynamical mass (\ref{mdyn}) remains finite, corresponding to a \emph{physical} fermion mass.
In the previous limit, the non-physical gauge field decouples from the theory, and we also note that the gauge dependence of the
dynamical mass is avoided.

An essential feature of the mechanism described here is the following. Although LIV features are suppressed by 
the large mass scale $M$, so that the corresponding effects are negligible at the classical level, quantum corrections completely change this picture, 
and lead to finite effects. In our present study, the finite effect is the dynamical generation of fermion masses, which is present 
even after setting the LIV-suppressing mass scale $M$ to infinity and the coupling $e$ to 0. 
Also, after this limit is taken, relativistic dispersion relations for 
fermions are recovered, such that the dynamical masses are the only finite effect from the original LIV model.

\section{Model}

The LIV Lagrangian considered in \cite{ALM} is
\be\label{Lag}
\mathcal{L}= -\frac{1}{4} F_{\mu\nu}(1-\frac{\Delta}{M^2})F^{\mu\nu}+\bar{\Psi}(i \slashed{\partial}-\tau \slashed{A})\Psi,
\ee
where $F_{\mu\nu}$ is the Abelian field strength for the gauge field $A^\mu$ and $\Delta=-\partial_i\partial^i$ is the Laplacian. 
The mass scale $M$ suppresses the LIV derivative operator $\Delta$, 
and can be thought of as the Plank mass, but which will eventually be set to infinity. $\Psi$ is a massless fermion doublet
$\Psi = (\psi_1,\psi_2)$ and the flavour mixing matrix $\tau$ features the gauge couplings $(e_1,e_2,\epsilon)$ as 
\be\label{tau}
\tau = \begin{pmatrix}e_1 & -i\epsilon \\i\epsilon & e_2\end{pmatrix}
=\frac{e_1 + e_2}{2}{\bf1} +\frac{e_1 - e_2}{2} \sigma_3 +\epsilon \sigma_2~,
\ee
where $\sigma_i$ are the usual Pauli matrices and ${\bf 1}$ is the $2\times2$ identity matrix. The fermions 
$\psi_1$ and $\psi_2$ in eq.(\ref{Lag}) are Dirac, but the structure of the gap equations which is discussed bellow remains the same in 
the case of Majorana fermions, hence  
the corresponding dynamical masses are independent of the nature of fermions. 

In order to study the possibility of generating masses dynamically, we assume the dressed fermion mass matrix
\be\label{Matrix}
{\bf M}=\begin{pmatrix} m_1 & \mu \\ \mu & m_2\end{pmatrix}
=\frac{m_1 + m_2}{2}{\bf1} + \frac{m_1 - m_2}{2} \sigma_3 + \mu \sigma_1 ~,
\ee
with eigenvalues $\lambda_{m\pm} = (m_1+m_2)/2 \pm\sqrt{(m_1-m_2)^2+4\mu^2}/2$. We allow for the presence of ${\bf M}$
in the gap equations, obtained from the self-consistent Schwinger-Dyson equation for the fermion propagator,
which has the usual structure and is not modified by the LIV term in the Lagrangian (\ref{Lag}). 
If we neglect corrections to the wave functions, the vertices and the
gauge propagator, the Schwinger-Dyson equation reads for our model 
\begin{eqnarray}\label{SD}
G^{-1}-S^{-1} = \int_p D_{\mu\nu}~ \tau\gamma^\mu ~G ~\tau\gamma^{\nu}~,
\end{eqnarray}
where $S,G,D_{\mu\nu}$ are the bare fermion propagator, the dressed fermion propagator and the gauge propagator respectively, 
given by (we denote $p_\mu=(\omega,\vec p)$)
\bea
S&=&i\slashed{p}/p^2\\
G&=&i\frac{p^2 + \slashed{p}( m_1 + m_2) + m_1 m_2 -\mu^2}{(p^2 - m_1^2)(p^2 - m_2^2)-2 \mu^2(p^2 + m_1 m_2)+\mu^4}\nn
&&\times\left[(\slashed{p} - \frac{m_1 + m_2}{2}) {\bf1} + \frac{m_1 - m_2}{2} \sigma_3 +\mu \sigma_1 \right]\nn
D_{\mu\nu}&=&-\frac{i}{1 +\vec{p}^2/M^2}\left(\frac{\eta_{\mu\nu}}{\omega^2-\vec{p}^2}+\zeta \frac{p_\mu p_\nu}{(\omega^2-\vec{p}^2)^2}\right)~,\nonumber
\eea
where $\zeta$ is a gauge fixing parameter.
The loop integral (\ref{SD}) is finite as a consequence of the LIV term $\vec p^2/M^2$ in the denominator of the gauge 
propagator.

\section{Solutions}

The Schwinger-Dyson equation (\ref{SD}) leads to 4 self-consistent gap equations, which 
must be satisfied by three unknowns $m_1,m_2,\mu$. This is possible provided constraints are satisfied, and
several possibilities are derived in \cite{ALM}, which are summarized here:

\nin{\bf(a)} \{$m_1=m_2=0$, $\mu\ne0$ and $e_1e_2>\epsilon^2$\} or \{$m_1=-m_2\ne0$ and $e_1=e_2>\epsilon$\}:
Mixing but no oscillation because the mass eigenvalues are opposite. In both cases we have
\be
m^2+\mu^2=M^2\exp\left(\frac{-16\pi^2}{(4+\zeta)(e_1e_2-\epsilon^2)}\right)~,
\ee
where $m=|m_1|=|m_2|$, and only the sum $m^2+\mu^2$ is constrained;\\

\nin{\bf(b)} \{$m_1=m_2\ne0$, $\mu^2=m_1m_2$ and $e_1=e_2$, $\epsilon=0$\}:
Mixing and flavour oscillations. In this situation, the mass matrix has eigenvalues
\be
\lambda_+=2m=M\exp\left(-\frac{8\pi^2}{(4+\zeta)e^2}\right)~~,~~\lambda_-=0~,
\ee
where $m=m_1=m_2$ and $e=e_1=e_2$. The mixing angle is $\pm\pi/4$, depending on the sign of $\mu$. Note that this mixing angle corresponds to $\theta_{23}$;\\

\nin{\bf(c)} \{$\mu=0$ and $\epsilon=0$\}: No mixing. In this case, the two eigenvalues of the mass matrix are
\be\label{2mass}
m_i=M\exp\left(\frac{-8\pi^2}{(4+\zeta)e_i^2}\right)~~~,~i=1,2~,
\ee 
so the mass matrix is diagonal in flavour space, with masses $m_i$.

The latter case is relevant to Majorana mass eigenstates. Indeed, one defines the Majorana fields $\nu_k$ in terms of the left-handed Dirac fields $\nu_k^L$ as 
$\nu_k=\nu_k^L+(\nu_k^L)^C$, such that the Lagrangian for free Majorana fields reads
\be
{\cal L}=\frac{1}{2}\sum_k\left(\ol\nu_k i\slashed\partial\nu_k-m_k\ol\nu_k\nu_k\right)
\ee
where the extra overall factor $1/2$ compared to the Dirac case does not play a role in our discussion, as it can be absorbed in a redefinition of masses.

\section{Discussion}

One can consider the coupling of a doublet of (mass eigenstate) Majorana fields to the regulator U(1) gauge field $A_\mu$ in the 
case {\bf (c)}. The fact that a Majorana field contains both chiralities allows for a straightforward extension of 
the Dirac case discussed in previous sections to the current situation. In this way, we are able to generate dynamically different mass 
eigenvalues (\ref{2mass}) for the two species, without mixing, as implied by the corresponding solution. This is a consistent way 
of discussing the dynamical appearance of a Majorana mass for left-handed neutrinos of the standard model. The non-trivial mixing of flavour 
eigenstates is then obtained by coupling neutrinos to the physical $SU(2)_L$ gauge fields of the standard model.
Thus it is because Majorana neutrinos are mass eigenstates that the solution {\it without mixing} is relevant to the scenario with Majorana neutrinos.

In order to recover Lorentz invariance in the different solutions found, we finally take the simultaneous limits
\be\label{limit}
M\to\infty~~~~\mbox{and}~~~~e_1,~e_2,~\epsilon~\to0,
\ee
in such a way that the dynamical masses are \emph{finite}, and we denote the corresponding ``renormalized'' mass matrix by ${\bf M}_R$. This
mass matrix can be fixed by experimental data, which in the interesting cases {\bf (b)} and {\bf (c)} leads to the couplings $e_1,e_2$ as functions of $M$, 
for the limit (\ref{limit}) to be defined unambiguously.
This procedure is independent of the gauge parameter $\zeta$, and the resulting
fermion mass is set to any desired value. 
In this limit, the gauge field decouples from fermions, and the only finite effect 
from Lorentz violation in the original model is the presence of finite dynamical masses for fermions. 
Indeed, the fermion dispersion relations are relativistic in the limit (\ref{limit}), since the fermion self-energy is then \cite{ALM}
\be
\Sigma(\omega,\vec p)~~\to~~-\frac{1}{4}(\omega\gamma^0-\vec p\cdot\vec\gamma){\bf 1}-{\bf M}_R~,
\ee
such that time and space derivatives are dressed with the same (non-perturbative) corrections.

We also note that, because of flavour mixing interactions of fermions with the vector field, that latter may become massive \cite{massiveA}, but
this feature does not occur here in the relevant cases {\bf (b)} or {\bf (c)}, where $\epsilon=0$.

Finally, the case {\bf (b)} might seem restrictive for Dirac fermions, since the mixing angle is necessarily equal 
to $\pm\pi/4$. This is a result of the high symmetry of the $2\times2$ mass matrix, and 
the extension to the three-flavour case is planned, in order to allow for more flexibility and phenomenological studies.


\vspace{1cm}

\begin{thebibliography}{99}

\bibitem{Higgs}
%\cite{Aad:2012tfa}
  G.~Aad {\it et al.}  [ATLAS Collaboration],
  %``Observation of a new particle in the search for the Standard Model Higgs boson with the ATLAS detector at the LHC,''
  Phys.\ Lett.\ B {\bf 716} (2012) 1
  [arXiv:1207.7214 [hep-ex]];
  %%CITATION = ARXIV:1207.7214;%%
  %897 citations counted in INSPIRE as of 10 Apr 2013
%\cite{Chatrchyan:2012ufa}
  S.~Chatrchyan {\it et al.}  [CMS Collaboration],
  %``Observation of a new boson at a mass of 125 GeV with the CMS experiment at the LHC,''
  Phys.\ Lett.\ B {\bf 716} (2012) 30
  [arXiv:1207.7235 [hep-ex]].
  %%CITATION = ARXIV:1207.7235;%%
  %888 citations counted in INSPIRE as of 10 Apr 2013
 
\bibitem{seesaw}
For a review see: R.~N.~Mohapatra, S.~Antusch, K.~S.~Babu, G.~Barenboim, M.~-C.~Chen, A.~de Gouvea, P.~de Holanda and B.~Dutta {\it et al.},
  %``Theory of neutrinos: A White paper,''
  Rept.\ Prog.\ Phys.\  {\bf 70} (2007) 1757
  [hep-ph/0510213] and references therein. 
  %%CITATION = HEP-PH/0510213;%%
  %328 citations counted in INSPIRE as of 22 Apr 2013
  
\bibitem{ALM}
%\cite{Alexandre:2013tya}
  J.~Alexandre, J.~Leite and N.~E.~Mavromatos,
  %``Lorentz-Violating Regulator Gauge Fields as the Origin of Dynamical Flavour Oscillations,''
  Phys.\ Rev.\ D {\bf 87} (2013) 125029
  [arXiv:1304.7706 [hep-ph]].
  %%CITATION = ARXIV:1304.7706;%%

  
\bibitem{JA}
%\cite{Alexandre:2010vh}
  J.~Alexandre,
  %``Dynamical mass generation in Lorentz-violating QED,''
  arXiv:1009.5834 [hep-ph];
  %%CITATION = ARXIV:1009.5834;%%
  %8 citations counted in INSPIRE as of 20 Mar 2013
%\cite{Alexandre:2011ux}
  J.~Alexandre and A.~Vergou,
  %``Properties of a consistent Lorentz-violating Abelian gauge theory,''
  Phys.\ Rev.\ D {\bf 83} (2011) 125008
  [arXiv:1103.2701 [hep-th]].
  %%CITATION = ARXIV:1103.2701;%%
  %8 citations counted in INSPIRE as of 20 Mar 2013
  
\bibitem{NM}
%\cite{Mavromatos:2010ar}
  N.~E.~Mavromatos,
  %``Quantum-Gravity Induced Lorentz Violation and Dynamical Mass Generation,''
  Phys.\ Rev.\ D {\bf 83} (2011) 025018
  [arXiv:1011.3528 [hep-ph]].
  %%CITATION = ARXIV:1011.3528;%%
  %10 citations counted in INSPIRE as of 20 Mar 2013
  
\bibitem{AFMP}
%\cite{Alexandre:2007na}
  J.~Alexandre, K.~Farakos, N.~E.~Mavromatos and P.~Pasipoularides,
  %``Neutrino oscillations in a stochastic model for space-time foam,''
  Phys.\ Rev.\ D {\bf 77} (2008) 105001
  [arXiv:0712.1779 [hep-ph]].
  %%CITATION = ARXIV:0712.1779;%%
  %9 citations counted in INSPIRE as of 22 May 2013
  
\bibitem{massiveA}
  %\cite{Jackiw:1973tr}
  R.~Jackiw, K.~Johnson and ,
  %``Dynamical Model of Spontaneously Broken Gauge Symmetries,''
  Phys.\ Rev.\ D {\bf 8} (1973) 2386;
  %%CITATION = PHRVA,D8,2386;%%
  %426 citations counted in INSPIRE as of 28 Mar 2013
%\cite{Cornwall:1973ts}
  J.~M.~Cornwall, R.~E.~Norton and ,
  %``Spontaneous Symmetry Breaking Without Scalar Mesons,''
  Phys.\ Rev.\ D {\bf 8} (1973) 3338.
  %%CITATION = PHRVA,D8,3338;%%
  %305 citations counted in INSPIRE as of 28 Mar 2013
  
  
\end{thebibliography}
\end{document}